# STUDENTS' CONCEPTS IN UNDERSTANDING OF SOUND

Zdeslav Hrepic

Department of Earth and Space Sciences, Columbus State University

4225 University Avenue, Columbus, GA 31907
drz@columbusstate.edu

**ABSTRACT**

The article describes alternative concepts in understanding of sound observed among students at the elementary, high school and college level. The research confirmed some previously found alternative concepts and pointed to several others. A majority of the difficulties were found to be common to all levels. Also the percentage of students, who express alternative concepts, is nearly the same at all educational levels. However the percentage of students who clearly don't have alternative concepts, in general increases with educational level.

One set of observed difficulties, as a whole, compose conceptually structured and coherent, but naive picture on propagation of sound. Data were collected through written open-ended test questions on sample of 287 pupils and students in Split, Croatia 1997.

**INTRODUCTION**

Teaching physics cannot be effective, in general, if a presentation does not take into account students' alternative conceptions. This statement is in agreement with recent findings in general science education studies[1,2]. Therefore instruction in physics should focus on students beliefs about the world, which means that such beliefs have to be identified[3]. To define important terms I use, I will present the list which according Dykstra at al[3] summarizes what different authors mean when they speak of alternative conceptions:

1. The "mistaken" answers students give when confronted with a particular situation: e.g. "The sun goes around earth."
2. The ideas about particular situations students have which evoke the "mistaken" answers. These are beliefs in a situated sense – belief about what will happen in a particular situation.
3. The fundamental belief students have about how the world works, which they apply to variety of different situations. These are beliefs in an explanatory sense about causality.

In later papers[4] this third category was, from cognitive perspective more appropriately, identified as a mental model - a robust and coherent knowledge element or strongly associated set of knowledge elements[5]. Therefore I will use term mental model in a sense described in third category above. For both, the first and second categories, I will use term alternative concepts as suggested by Wandersee at al[6]. Spontaneous mental models and alternative concepts are generally different from scientific models and may contain contradictory elements[4]. Models are scientifically accepted as valid if are found to be coherent, stable and experimentally approved. During the teaching process we want, in a sense, to "replace" alternative concepts with those that are scientifically accepted. Therefore, of significant importance is also the frequency and origin of students' alternative conceptions.

The aim of this research[7] was to investigate students' alternative concepts related to the sound, to find their representation at different educational levels and possibly to identify their origins.

## RESEARCH METHODS

Data were collected using a written survey with open-ended, mostly original questions. (See the Appendix for a listing of all test questions).

Some questions (A6, A10, B5, C9, C10, C11 and D3) were adapted from Hewitt's "Conceptual Physics"[8], and some were taken from Pereljman's "Interesting Physics"[9] (A1, A2, and C7). It seems that Aristotle was first to be concerned with question D10.

Testing was completed during 1997, immediately after students had finished classes on sound. The examinees were 8th graders in two Elementary schools, juniors in three high schools with different emphasis (on mathematics, modern languages, and electro engineering), and seniors at two Colleges (The College of Technology and The College of Natural Sciences, Mathematics and Education), all in Split, Croatia. None of the teachers, whose classes took the test, were acquainted with the test while they were teaching the sound. All test questions (44 altogether) that have been given divided into four test groups, are appendixed at the end of this paper in same format in which they were given to high school and college students. Elementary school pupils answered 28 (of these 44) questions, also divided in four groups.

## RESULTS

The fundamental result of this research is clear isolation of a number of alternative concepts. Their most probable origins have several common denominators and alternative concepts are grouped here accordingly. The idea about sound propagating as a particle was found to be stable, coherent, and a widely applicable alternative mental model of sound propagation at all educational levels.

## 1. Particle mental model - first "law" of spontaneous acoustics
1.1. Sound propagates as a particle-like object

## 2. Alternative concepts about propagation of sound generated by the particle model
2.1. Material obstacles slow down propagation of sound
2.2. If louder, sound travels faster
2.3. The speed of sound depends on movement of the sound source
2.4. Sound can be perceived in distance like a distant object

## 3. Alternative concepts generated by inappropriate knowledge transfer
3.1. Not all the materials can propagate sound
3.2. Sound energy is not generally transformable

## 4. Alternative concepts generated or enforced by school knowledge
4.1. The denser the medium, the faster sound propagates
4.2. The speed of sound depends on its frequency
4.3. Wind influences the frequency of received sound

A table I. below shows the extent to which these concepts were represented. The percentages are based on only one question, in which the respective model was most obvious. Status of alternative concepts in the table I. is divided into "Students who have it" and "Students who do not have it" column. The first one represents those examinees whose answer affirms respective alternative concept and the second one those whose answer negates it. The former, negation group, besides fully correct answers also includes correct answers with partially correct explanation or without any explanation (although it was required in all questions). This way the statistics deals with concepts' presence rather than with correctness of answers.

The open-ended responses resulted in variety of different answers, and for one part of them I couldn't claim they belonged to any of two stated groups. These answers are in column "others with incorrect or no answer". Typical answers in this group are not answered questions, answers with missed topics, correct statements but with completely wrong explanations and similar.

Table I: Alternative concepts and their statistical representation.

|  | Alternative concept status | | | | | | | | |
|---|---|---|---|---|---|---|---|---|---|
| Alternative Concept/Model | Students who have it | | | | Students who do not have it | | | | Others with incorrect or no answer |
|  | Elementary School | High school | Colleges | Together | Elementary School | High school | Colleges | Together |  |
| 1.1. Sound propagates as a particle-like object | N.A. | N.A. | N.A. | N.A. | N.A. | N.A. | N.A. | N.A. | N.A. |
| 2.1. Material obstacles slow down propagation of sound | 64% 18/28 | 49% 21/43 | 100% 8/8 | 60% 47/79 | 11% 3/28 | 28% 12/43 | 0% 0/8 | 19% 15/79 | 22% 17/79 |
| 2.2. If louder, sound travels faster | 55% 11/20 | 29% 12/41 | 17% 1/6 | 36% 24/67 | 40% 8/20 | 66% 27/41 | 83% 5/6 | 60% 40/67 | 4% 3/67 |
| 2.3. The speed of sound depends on movement of the sound source | 71% 20/28 | 52% 15/29 | 63% 5/8 | 62% 40/65 | 7% 2/28 | 35% 10/29 | 38% 3/8 | 23% 15/65 | 15% 10/65 |
| 2.4. Sound can be perceived in distance like a distant object | N.A. | N.A. | N.A. | N.A. | N.A. | N.A. | N.A. | N.A. | N.A. |
| 3.1. Not all the materials can propagate sound | 64% 18/28 | 35% 15/43 | 25% 2/8 | 44% 35/79 | 21% 6/28 | 56% 24/43 | 21% 6/8 | 46% 36/79 | 10% 8/79 |
| 3.2. Sound energy is not generally transformable | 85% 17/20 | 62% 23/37 | 50% 3/6 | 68% 43/63 | 10% 2/20 | 11% 4/37 | 33% 2/6 | 13% 8/63 | 19% 12/63 |
| 4.1. The denser the medium, the faster sound propagates | 32% 8/25 | 55% 17/31 | 50% 4/8 | 45% 29/64 | 40% 10/25 | 29% 9/31 | 50% 4/8 | 36% 23/64 | 19% 12/64 |
| 4.2. The speed of sound depends on its frequency | 46% 13/28 | 51% 22/43 | 25% 2/8 | 47% 37/79 | 21% 6/28 | 42% 18/43 | 63% 5/8 | 37% 29/79 | 16% 13/79 |
| 4.3. Wind influences the frequency of received sound | N.A. | 29% 9/31 | 50% 4/8 | 33% 13/39 | N.A. | 13% 4/31 | 0% 0/8 | 10% 4/39 | 56% 22/39 |
| Together | 59% 105/177 | 47% 134/298 | 48% 29/60 | 50% 268/535 | 21% 37/177 | 36% 108/298 | 42% 25/60 | 32% 170/535 | 17% 93/535 |

# DISCUSSION - ELABORATION OF ALTERNATIVE CONCEPTS

In each elaboration of alternative concept I will give the accepted scientific model, more detailed description of the alternative concept where necessary, and a review of a previous work related to concepts that were already described. Also, after stating or referring to the question that was the basis for statistics in the Table I., I quote a certain number of characteristic students' answers, which describe a respective concept. Before every answer, the group and ordinal number of the question is denoted (see tests in appendices), together with the abbreviation for school level of student who gave the answer. Abbreviations for school level are: (El.)-for Elementary schools, (H.S.)–for High Schools and (Col.)–for Colleges. Responses have been translated from Croatian as close to original sentence as possible. Where necessary, I have added short italic inserts *[within square parentheses]* into answers so that referring to the questions is not necessary to follow the meaning.

## 1. Particle mental model - first "law" of spontaneous acoustics

Two alternative models of sound propagation were identified. From this research it seems they are not only most significant problem of "spontaneous acoustics", but also basis for several other alternative concepts. The first one is "particle model", which was noticed, and in different ways described by Linder & Erickson[10], Linder[11], Maurines[12] and Barman, Barman & Miller[13]. In this model the sound propagation is perceived as travel of an "entity"[11], "bounded substance"[11] or "sound particle"[12] through a medium. The other model involves particle pulses model described by Wittmann, Steinberg & Redish[14,15]. It represents a "pulsating" variation of previous one e.g. students perceive sound propagation as translational movement of sets of particles emerging consecutively from the source and propagating away from it. Both models support Wittmann's finding that in the context of wave physics students often reason by focusing solely on object-like properties of the system they consider[16]. This "materialistic conception" seems to be very fundamental alternative concept in physics as with matter-like properties students endow also things such as electricity, light, and heat[17].

### 1.1. Sound propagates as a particle-like object

A great number of students' answers revealed the concept of propagation of the sound consistent with "particle model" described above: sound propagates like the multitude of tiny material entities. These particle-like objects have momentum, volume and even shape and are sent off spherically from the source in all directions. These objects have been referred to as sound particles[12,14]. They propagate in a translation manner - from starting point to destination through the obstacles on the way. There was no specific, particular question directly related to this model, so I cannot unambiguously give its statistical weight. However its existence can be clearly seen in many independent answers on different questions and, as described later, through the set of alternative concepts that seem to be direct outcomes of this model.

Let me start with two statements where particles of sound were actually mentioned:

B2 (H.S.): "*[Under the water]* we hear sound *[coming from the air]* weakly and unrecognizably because the particles of the sound can not penetrate the water. It has a dense molecular structure"

A9 (H.S.): "*[The sound can do the work]*, because the particles of the sound are hitting the obstacle..."

As any other material object, while passing through a medium, the sound "collides" and "acts with force" on material particles, which stand in its way:

A8 (Col.): "The less the density, the greater the propagating velocity...it is logical that if a wave collides with particles, the propagating velocity will be smaller."

B2 (H.S.): "Sound propagates by acting with a force on the particles of material through which it passes. The denser the material – the less sound propagates..."

When an insulator is involved:

A5 (E.H.): "...well, sound will pass through an insulator only if there are some holes, cracks through which it can creep through."

In coherence with the particle concept is certain aerodynamics of sound (observed only in elementary schools):

C8 (El.): "I think that the sound of a violin, which has stronger *[literal translation]* frequency travels faster, because its sound is thinner and faster in the air."

C8 (El.): "*[Sound of violin] t*ravels faster *[than sound of contrabass]*. Because violin sound is sharper."

As a material body, the sound occupies certain volume and it will not be able to pass if there is no enough space:

A8 (El.): "If the density is greater, sound creeps through harder, because there is no space for it. Otherwise we would hear sound immediately."

A8 (El.): "Sound passes slower *[in denser materials]* because it is dense and there is no space for it."

C2 (El.): "It is not possible *[for honey to propagate the sound]*. Honey will fill all cavities, every cell in the air."

## 2. Alternative concepts about propagation of sound generated by the particle model

My hypothesis is that the previously described mental model and corresponding generalizations from the mechanics of moving bodies is the origin of whole this set of alternative concepts.

### 2.1. Material obstacles slow down propagation of sound

According to this alternative concept the sound is slowed down when it comes across objects denser than the medium in which it propagates. This is consistent with the particle model of sound propagation. Sound is "breaking through" a medium, and the particles/density of medium make propagation difficult.

In 1993 two authors reported students' difficulty related to dependence of density of the medium and velocity of the sound in it. Maurines[12] writes that both, students who did and who did not receive lessons about waves, most often state that "the propagation is more difficult when the medium is dense". This is correct. Speed of sound is inversely proportional to square root of density of the medium, but forces between the particles of the medium play equally an important role in determining the speed of sound. Although correct, the idea about reverse proportionality of density of medium and speed of sound leads to an false conclusion: "Whereas most of the students answer that the sound velocities depend on the medium, they classify them in the reverse order: sound propagates more quickly in vacuum than in water or in steel[12]."

Also in 1993, Linder[18] presented a more comprehensive definition of this concept: "The speed of the sound is a function of the physical obstruction that molecules present to sound as it navigates its way through the medium" and also pointed out the source of the concept: "conceptualized as a physical thing, sound is slowed down by physical obstacles…[18]"

In my research, this alternative concept was most clearly expressed in responses to question C3 (N=79), in which the (wooden) barrier is very obvious (please refer to the question 3 in group C of the appendix). 60% of respondents explicitly state the barrier slows down the sound and therefore the sound will be heard first by listener B. 19% correctly stated the sound would propagate faster through the wood than through the air. 18% state that both listeners will hear the sound at the same time (all percentages that describe the representation of particular answer within respective question are relative to total number of students answering respective question).

Characteristic "alternative" answers are practically the same at all levels:
C3 (El.): "Sound will arrive first to the listener B because there is a wooden barrier on the other side. It is because sound propagates slower if it comes across the barriers."
C3 (H.S.): "To listener B sound will arrive first, because the obstacle presents a barrier and sound will slowly come to listener A."
C3 (Col.): "To listener B, because on the way to A there is wooden barrier which slows down speed of sound."

The answer below is an example of statements that nicely points out the background of this concept - the previously described particle model of propagation of the sound.
A8 (El.): "If particles *[of a medium]* are denser, sound will move more slowly because it must break through this density."

We all know from common experience that material barriers obstruct the movement of physical objects. Denser ones generally do more. Moving through the water is more difficult than through the air. Likewise, when we reach the wall we cannot go further. So, perceived as a material object, the sound obeys the same rules.

**2.2. If louder, sound travels faster**
Maurines described this alternative conception[12] but the reasoning about proportional relation of signal amplitude and its speed was observed also in the cases of other waves and pulse-like signals[12,14,19]. Probably closely related to this concept that louder sound travels faster is also a broad belief that the speed of wave propagation depends on the force used to create a wave[19,15]. I share belief of some other authors[19] that this reasoning comes from simple analogy with setting a mechanical object into motion: a more energetic hit, a more powerful motor or a stronger swing give greater speed to a body. E.g. a ball flies faster if we throw it harder. But, being a mechanical wave, in an ideally isotropic medium sound has the same speed regardless of its amplitude and intensity. Here, this alternative concept appeared in question D3 (N=67): Does sound of strong loudspeaker travel faster, slower or with the same velocity as sound of weak loudspeaker if they are both on maximum? Why?

36% of all examinees share the alternative view that louder sound travels faster. 60% hold correctly that speed of sound does not depend on its loudness. 40% gave correct explanations, too: the speed of sound does depend only on medium, or, the speed of sound does not depend on intensity/loudness.

Although the frequency was not mentioned in this question at all, 22% stated that the frequency of the sound is the only factor in determinining the answer (see concept 4.2). Another 10% related sound of the stronger loudspeaker to higher frequency of sound ("…stronger loudspeaker produces sound vibrations at faster rate").

Characteristic answers are:

D1 (El.): "Sound of the stronger loudspeaker travels faster than the sound of the weaker one. It travels faster because it has stronger power and the sound automatically spreads faster."

D1 (H.S.): " The sound of stronger loudspeaker travels faster because it is stronger."

D1 (El.): "The sound of stronger *[loudspeaker]* vibrates faster and therefore travels faster"

**2.3. The speed of sound depends on movement of the sound source**

Concept that speed of propagating entity depends on movement of its source/launcher is another one, which is false for sound (waves), but correct for material objects. For this reason I believe the mechanics of moving bodies is the background of this concept too. As an example, the velocity of man walking on the moving train (with respect to ground) depends on velocity of the train. It is true also for velocity of the bullet fired from train, which would be an even closer example for this transfer of reasoning. It is also possible that this concept leans on previously described idea that more energetic/forceful source creates faster signal/wave/sound, and fast source is in a way - more energetic. The cornerstone question for this concept was C10 (N=65) : Does sound of car siren travel equally fast when the car is at rest and when it is moving towards the observer?

The alternative concept shown through this question has two "stages". The general one is that the speed of the sound (with respect to ground) depends on movement of the source. It was expressed by 62% of examinees (see Table I). Another stage is a dominant "sub concept". It states that the speed of sound emitted from a moving source increases in the direction in which the source moves. This specific concept was expressed in 43% of all answers. Stated reasons were: "frequency gets higher", "distance gets shorter" and "the speed of automobile carries the sound". Characteristic answers are:

C10 (El.): "The sound of siren spreads faster if the automobile moves because the automobile (although with small speed when compared with sound), still does move..."

C10 (H.S.): "It travels faster when the automobile moves because besides the speed of the sound there is speed of the automobile which carries that sound."

C10 (H.S.): "Speed of sound towards listener is increased by the amount of the speed of the automobile."

**2.4. Sound can be perceived in distance like a distant object**

The sound produces sensation of hearing upon reaching our ear. It seems that several students think it can be heard also when it is far from the observer.
I have noticed this alternative concept only in several answers (no statistical data applicable), so I will quote two answers, which clearly expose it (to see question D11 please take a look at the appendix).
D11 (Col.): "Yes, *[people will hear each other across the high wall]* because if they move about 10 m from that wall and if they speak a little louder, the sound wave will spread around. That part of sound wave at the wall level, the other man will not hear because the wall will not let the sound wave to pass through, but the rest of sound wave which passes above the wall, man will hear as if he listens to it on the much greater distance."
So, because the barrier prevents the sound from reaching the listener's level, it will be passing above the wall, and thus also above the listener, that way not reaching his level. Nevertheless, the listener will hear it - as if being at "the much greater distance".
The same alternative concept was revealed in couple of answers in question A6: Suddenly an airplane, which flies at ultrasonic speed, appears right above the listener. What can the listener hear from the plane? Why?
A6 (H.S.): "As the *[ultrasonic]* plane is faster than sound, the plane will reach the listener first, and sound will reach him later. This means, at this moment *[when the plane is right above him]*, listener will hear distant sound of the plane..."
The student is aware of the fact that the sound could not reach the listener at the moment when ultrasonic plane was above him ("it will reach him later"). So this "distant sound" that will be heard, is a sound that did not reach the listener.

I hypothesize the background of this concept, is commonly seen situation that a distant object, and also a wave on the water, can be easily noticed even when they are far from observer. Problem appears when it is not taken into consideration that light waves in those cases really do reach us.

# 3. Alternative concepts generated by bad knowledge transfer

Next three concepts are most likely consequences of poor understanding of the role of propagating medium and relation of the sound with the other physical phenomena (in this cases electricity, energy and light).

**3.1. Not all the materials can propagate sound**

Sound waves propagate through any material medium although speed and attenuation may be very different in different materials. Nevertheless, the ability of some materials to propagate sound (in principle) seems to be doubted by relatively large number of students. Of these suspicious materials I have probed honey and plastics. Question: "Does plastic propagate sound?", I probed in the conversation with my acquaintances before I have put it on test. Some interesting answers were:
Senior in physics and polytechnics: "It should not."

The teacher of physics and polytechnics: "It does not propagate, I suppose..."
The teacher of mathematics and physics: "It seems to me that plastic does not propagate anything, neither electricity, nor damp, nor sound."

In the answers on the same test question, (Question D6, N=67), 31% stated plastic doesn't propagate sound, 9% stated that it propagates poorly (because of big stiffness, big density, its structure, and its insulating characteristics) and 45% stated that it does propagate (because of its structure/density).

Reasoning about propagating characteristics of the plastics in few students' answers, indicate analogy with electrical conductivity of the material.
D6 (H.S.): "No, plastic does not propagate sound because it is not a conductor."
D6 (H.S.): "It depends on thickness of plastic and on its insulating characteristics, that is on resistance to passage of wave.
D6 (Col.): "Propagates, but very poorly because it has a very small number of free particles which can vibrate."

The honey as a medium was questioned more. When asked if the honey propagates sound (Question C2, N=79), 44% stated it doesn't, and 15%, that it does but poorly. Most common explanation was – because of a big density. 30% said the honey does propagate the sound (because it has particles, density). The data in table I. related to this concept is based on question C2. Characteristic answers on C2 were.
C2 (H.S.): "It does not *[propagate sound]*. Honey is too dense medium."
C2 (H.S.): "… because of the density of the honey, the sound moves slower there, if it can at all.
C2 (Col.): "Honey does not propagate sound, because its density is big enough to prevent propagation of sound."
We can here again see previously described "particle" and "density – barrier" models.

**3.2. Sound energy is not generally transformable**

Sound and light are two different forms of the energy and as such are convertible. According to this alternative concept the sound cannot be transformed into light by any means. The concept showed up in question B7 (N=63): Can we in some way transform sound into light? Explain why.

13% of examinees do allow for the possibility of this transformation. Three of these (5%) offered explanation for their statement but only one was complete and correct: "It can, because sound is also form of energy and energy can be transformed from one form into another." Interestingly, an elementary school pupil gave this answer.

Two other explanations can be said to be not incorrect: We can transform it "with appropriate technological solution"; and: "while breaking the sound barrier".

68% of examinees hold that this transformation is not possible. The reasons against this transformation are very different. Here are some examples:
B7 (El.): "No it can not. These are two completely different phenomenon."
B7 (H.S.): "It can not, because light is transversal, and sound is longitudinal wave."
B7 (H.S.): "No, because such a transformation of energy is not possible."
B7 (H.S.): "It can not, because sound is movement of matter while light represents quantum of electromagnetic radiation."

B7 (H.S.): "Sound can not be transformed into light, because sound is mechanical wave, and light is electromagnetic, so that their natures are completely different and incompatible."
B7 (H.S.): "No, because sound has not speed the of light."
B7 (H.S.): "No, because the sound is something perceived with ear and the light is perceived visually.
B7 (Col.): "No, because sound is mechanical wave, and light is electromagnetic."

So sound and light seem to be perceived as two inherently different phenomena. Also an idea exists that only "similar", or in a way, "compatible" "things" are transformable one into another.
A whole set of additional reasons for "incompatibility" of the sound and the light can be found in a some previous reports[11,20] related to this topic: "light has different sources than sound"; "the light will propagate through certain things the sound won't"; "sound bends…but light… does not"; "they are different kinds of energy"; "Doppler effect is commonly seen in the case of the sound but not light"; and so on.

## 4. Alternative concepts generated or enforced by school knowledge

Observed difficulties that are related to school knowledge can be grouped into three major sets. First two are misuses of terminology (e.g. "These two people should have frequency strong enough in order to hear each other…"), and misuses of the formulas ("the sound of the violin travels faster *[than sound of contrabass]*, because of higher frequency, meaning it needs less time. $\nu=1/T$". Although relatively frequent, these kinds of difficulties are very fragmented and without clear pattern. The last group of school-generated difficulties are alternative concepts and students' reasoning patterns related to them is far more structured and uniform that of first two.

### 4.1. The denser the medium, the faster sound propagates

Speed of the sound is inversely proportional to the square root of the density of the medium, which it propagates through. The observed alternative concept states the opposite. Following answer on the question B4 nicely describes the concept and points out its source:
B4 (H.S.): "The sound is a wave. It vibrates through any material. For example, the speed of the sound through the air is approximately 340 m/s, the speed through water is 1500 m/s, the speed through steel is 5000 m/s, which means that when the particles of the material are denser the waves of the sound move faster from a particle to particle respectively conveying energy..."

Linder[18] described basically this same concept this way: "The speed of sound is a function of molecular separation". His interviewees depict the model in which sound is an entity carried by molecules of medium for a certain distance and then transferred to other ongoing molecules. If molecules are closer (therefore medium denser) "they wouldn't have to travel as far to push one another" and consequently the speed of sound will be greater in denser materials.

In this research, the question A8 (N=64) very often revealed the same final conclusion but just occasionally the reasoning behind the answer was similar to this described by Linder, as in following examples:

A8 (H.S.):" For sound (mechanical wave) the medium is necessary. If it is necessary that means that if the material is denser the speed of the sound is greater."

A8 (Col.):" When the density is greater the sound is faster because the particles are nearer so the activity among them is conveyed faster."

Most often the background of the conclusion that the speed of the sound is proportional to density was students' familiarity with the actual speeds of the sound in some common materials like air, water and steel, which are given in comparison tables in practically all textbooks. When comparing densities of the materials with these speeds only, one easily comes to wrong conclusion about direct proportionality of these two quantities.

Students' reasoning about this was surveyed through question A8 (N=64): How does the density of some matter influence the speed of sound in it? Explain.

45% examinees stated the speed of the sound in material is directly proportional to the density, and 36%, correctly, that this relationship is inversely proportional. Examples of characteristics answers of proportional relationship:

A8 (El.): "I've heard that sound spreads faster in water than in air, so if water has greater density it means that if the density of the material is greater then the sound spreads faster."

A8 (H.S.): "When the material is denser the sound spreads faster. An example: "The sound is faster in water than in air""

A8 (H.S.): "The density of some material influences the speed, the sound moves faster through the material which is denser. It moves the fastest through steel.

On the other hand, of those 36% of examinees who (correctly) stated the inversely proportional relationship of speed and density (who are under "not having alternative concept"), none explained the answer in terms of the accepted scientific model. 14% offered explanation in terms of previously described particle/obstacle concepts of propagation of sound e.g.:

A8 (El.): "Bigger the density of material, breaking through is more difficult for sound."

One explanation was in terms of "resistance" and only 6% (4 of 64) offered an explanation similar to a partial scientific model. 14% did not give any explanation.

It is interesting to notice that this alternative concept is in direct confrontation with one previously stated about obstacles slowing down the sound. In previous concept (2.1.), we had 60% versus 19% of answers in favor of statement that material obstacles (denser materials) - slow down propagation of sound. Here we have 45% versus 36% in favor of the statement that denser medium propagate sound faster. Which reasoning will be triggered depends a lot on how a question was formulated (context) and on formal knowledge of students (familiarity with speeds of sound in different mediums). In this research, the alternative concept that the density (the obstacle) slows down the sound was prevailing in elementary schools. In the high schools the opposite one prevailed - the density speeds up the sound. The college students' opinions were rather equally divided. But as we saw - not having one of these alternative concepts very likely just means having another one, and not having the scientific model.

In the end a question of how to overcome this problem raised. It can be effectively done using the model of the masses hanging from the common horizontal support and connected with the springs. If we set mass at the one end of apparatus in the motion, the students usually notice easily that speed of propagation of the disturbance depends on several factors. If hanging masses are bigger or if there is greater number of masses at given length – the speed of the signal will be slowed down. These two factors are related to density of the "medium". Also, the stronger connections among these masses will increase this speed, which is related to elasticity of the medium.

**4.2. The speed of sound depends on its frequency**

In an ideally isotropic medium, the speed of sound does not depend on frequency. The test question related to this concept was C8 (N=79): Does the sound of violin travel faster, slower or with the same speed as the sound of contrabass, if we know that the violin produces higher frequencies. Explain.

In response there was 47% of the answers stating that the frequency of the sound does affect its speed. As a subconcept, 37% further declared this relationship to be proportional. Another 37% of the answers were correct (speed is the same) but only 15% also gave a correct explanation ("speed of sound depends only on the medium" and "speed of sound does not depend on frequency"). Conviction that frequency influences the speed of the sound mostly comes out from the wrong interpretation of the formulae $v=\lambda f$, from which students often conclude that speed of wave (v) proportionally increases with its frequency (f).

Following students' answers illustrate this reasoning:
C8 (H.S.):" The violin makes sounds of greater frequency than the contrabass does and as we know that the frequency and speed are proportional, the speed of the violin sound will be greater than the speed of the sound of the contrabass."
C8 (H.S.): " $v=\lambda/T=\lambda f$ . The sound of the violin travels faster."
C8 (Col.): "The sound of the violin travels faster because the speed depends on frequency; so the higher is the frequency, the greater is the speed."

Pupils in the elementary school often applied the "particle" nature of the sound in their answers (although they are also familiar with relation $v=\lambda f$). As an example:
C8 (El.): " The sound of the violin travels faster because it is thinner; it produces more vibrations per second. I would describe it as the lighter sound, because the sound of the contrabass is deep, it spreads across the air very hardly."

So, acquaintance with the relation $v=\lambda f$, especially among higher-level students, frequently generates and/or confirms the idea of the proportional dependence of frequency and the speed of the sound. It would be interesting to see if students would apply the same principle to the relation of wavelength and the speed of the sound.

**4.3. Wind influences the frequency of received sound**

Constant wind does not influence the frequency of received sound although it changes its speed, because the wavelength also changes and compensates for it (the ratio $v/\lambda$ remains unchanged). An alternative concept states opposite and was observed through question A10 (N=39). To see the question please refer to appendix.

The general form of the alternative conception is: Wind changes the frequency of the sound that a listener receives and its statistics is given in the table I. 33% examinees express this concept. Another level is again dominant "sub concept": The wind blowing from the source toward listener increases observed frequency. The wind from the opposite direction decreases it. 23% of examinees stated this concept in this specific form.

This specific form of concept but in terms of "speed" (instead the frequency of the sound), was expressed by another 23% of examinees, and further 7% expressed it in terms of "power". These statements are correct but none of these two terms was mentioned in the question. I do not have good explanation why they appear in answers.

Only 10% of the students have correctly concluded that the wind will not influence the frequency of the sound. 3 out of these four responses were without explanations and just one correctly stated:
A10 (H.S.) "It *[the wind]* will influence only the speed of the sound, but not its frequency."

The typical "alternative concept answer" is:
A10 (H.S.):" Yes, it will. The frequency will be higher if the speed of the sound is greater and vice versa.
A10 (Col.): "…Yes, it will. We have a feeling that the source was displaced and according to Doppler's effect it influences the frequency.
A10 (Col): "…$v = \lambda/T$, $f = 1/T$, $v = \lambda f$ → $v \sim f$"

It seems that one source of this concept is previously described alternative concept. Namely, if the wind increases the speed of the sound, which is true (and fits both - wave and particle model of propagation), and the speed is considered being simply proportional to frequency – the result is this new concept.

## CONCLUSIONS

This research has shown a number of alternative concepts related to students understanding of the sound. Some were newly found, and some which have been previously reported[21] are confirmed. The indications about their origin were found for most of them. Overall percentage of those who do not express the alternative concepts is 32% and, as expected, rises with level but probably not as rapidly as we might think or hope: 21% (El.) - 36% (H.S.) and 42% (Col). Another problem is relatively big percentage of students who do express alternative concepts (overall is 50%) and this percentage varies only slightly with level 59% (El.) - 47% (H.S.) - 48% (Col.).

Further, from this and other previously mentioned research, a general categorization of students difficulties related to sound could be drawn out in terms of their sources:
1. Students' models of propagation of the sound and their consequences.
2. Relation of the sound to the other physical phenomena (waves, light, energy, electricity, pressure, force…)
3. Misinterpretations of graphics, formulas, definitions and terms learned in school.

Although listing the concepts and finding their origins, is necessary as a starting point, crucial question to be addressed is – what is next step in overcoming identified alternative concepts? There is a need to translate existing alternative concepts research into practical lessons and activities that teachers can use in their classrooms[6] and recent articles on acoustics show this tendency[22,23]. Maybe best example is Wittmann's tutorial[15], constructed on the basics of his findings[14], which attempts to change student's perception of wave as an object to perception of wave as an event, or series of events.

However, when the sound is concerned, there is an unfortunate situation that, when compared with other fields, the number of reported researches in this domain is scanty. Therefore a remark that there is need for theoretical foundation that can describe, predict and explain alternative concepts[6] is especially valid here. In searching for underlying principles that governs sets of observed difficulties and alternative concepts one of promising approaches seems to be representation of students' reasoning in terms of David Hammer's resources[24]. He sees student's conclusions about physical phenomena as outcomes of resources they bring to learning. Alternative conception appears after inappropriate resource is activated. He sees Wittmann's tutorial as an attempt to replace the activation of an inappropriate resource – "sound is an object" - with an appropriate one – "sound is an event". In this sense the particle model of sound propagation and alternative concepts related to it, as described here, can be viewed as a consequence of activating the resource "wave is an object" or "sound is an object". Similarly described alternative concepts related to school knowledge are triggered by resources learned in school. When the relation of the sound to the other physical phenomena is concerned it seems that a variety of different resources are involved. With this approach in mind, after identifying students' alternative conceptions, instead of trying to "replace" them, instructor should identify the resources that activate these conceptions and the resources that activate the scientific model. Then the instruction can be designed to help students use their resources in a productive way.

## ACKNOWLEDGMENTS:

I express my deepest gratitude to all teachers and professors who have helped me to carry out this research. Besides, my special appreciation goes to Professor Ivica Luketin for his generous and friendly help and a lot of time he devoted to me.

[24] David Hammer, "Student resources for learning introductory physics," American Journal of Physics, Physics Education Research Supplement **68**, S52-S59 (2000).

**APPENDIX:**
**TEST QUESTIONS** - According the groups in which they were given
Note: There was an written instruction one at the beginning of each test:
Please explain every answer logically. You will do that best if you write your opinion on paper as if you were trying to convince your friend in consistency of your opinion. Please answer all questions.

# GROUP A

1. We hear higher sound when a mosquito flies than when a bumblebee flies. Why?
2. Will echo be heard in situations shown on pictures a and b?
Explain. (In both cases man looks in the direction of the arrow)

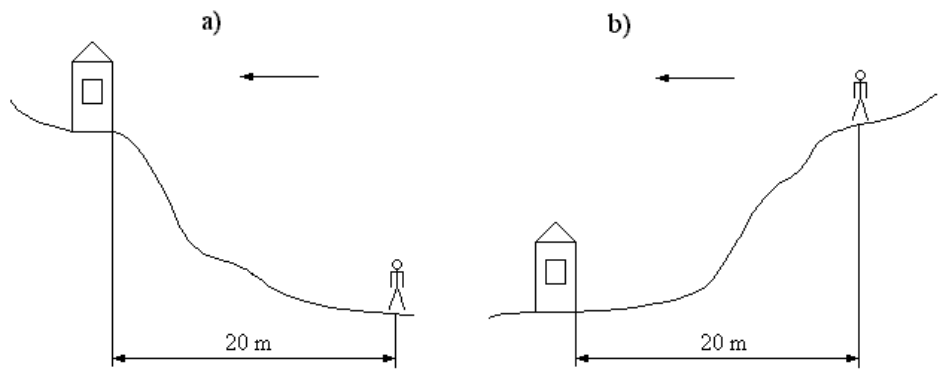

3. On what depends how high the tone made by whistling will be? In what way? Why?
4. Can we hear the signal from telephone earpiece if we turn it in reverse direction and put it on our ear? Why?
5. a) Explain the term sound insulation. What does a good sound insulator "do" with the sound?
   b) Can sound be absolutely insulated in some space? Explain.
6. Suddenly, an airplane, which flies at ultrasonic speed, appears right above the listener. What the listener can hear from the plane? Why?
7. If we hit a thin glass with a nail, we hear a clear sound. Why is this sound weaker if we hold the glass with one hand and hit again?
8. How does the density of some matter influence the speed of sound in it? Explain.
9. Is the sound capable of doing work? Explain.
10. A source of the sound with constant frequency (pitch) is placed in front of the observer at the distance of approximately 15 m (in open space). If the wind starts to blow from the source of the sound towards the observer, will it and how will it influence the frequency of received sound? What will happen (with the received frequency) if the wind blows from the opposite direction?
11. If a speaker speaks in a funnel (like this in the picture) and directs its bigger hole toward the listeners, they will hear louder sound than without the funnel. Why?

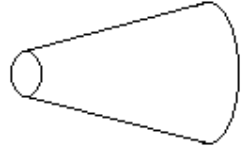

# GROUP B

1. Why we can hear sound from a big shell when we lean it on our ear?
2. a) Could we hear music from the radio placed in the bathroom if we dive our head into the bathing-tub filled with water?
   b) If this radio is water resistant, and if we dive it into this tub together with our head, could we also hear it?
   c) If only the radio is dived, could we hear it from outside? Explain why.
3. Why does the guitar string vibrate with higher frequency when pressed, than when it is free?
4. Can the sound be stopped? Explain.
5. Observer sees an airplane in one part of the sky, and hears its sound from the other (behind the plane). Does the airplane fly with the speed, which is less, greater or equal to the speed of the sound? Explain your answer.
6. Do clouds slow down the sound? Explain.
7. Can we in some way transform the sound into light? Explain why.
8. Two door wings are made of the same wood but of different thickness. Wideness and altitude of both are equal. Will inmates hear louder knocking if we identically knock on the thinner or on the thicker wing? Why?
9. Could sounds of two violins, playing simultaneously, be cancelled, so that we don't hear any of them?
10. How the echo originates and why?
11. The picture below shows devices made of metal discs on wooden pedestal. In the middle of the disc A there is a small ballot hanged on a thin rope. In experiment 1, discs A and B have the same dimensions. If we hit the disc B with hammer, the ballot on the disc A will vibrate too. Will vibrating of the ballot be greater, smaller or the same if we increase the diameter of disc A (thickness remains the same) and hit B again equally as before (experiment 2)? What if we decrease the diameter of disc A (in both experiments pedestal neglectedly decrease vibrating). Explain.

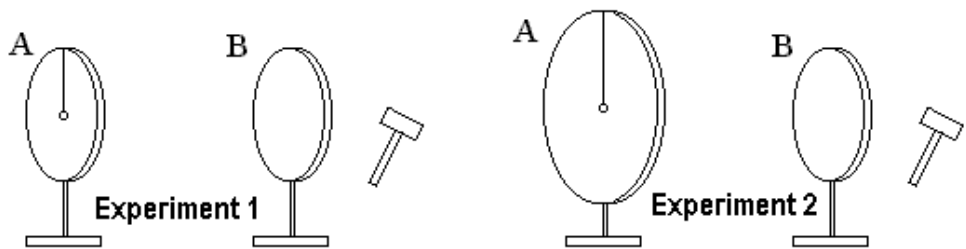

# GROUP C

1. Could we hear the sound of the Moon rotation if it would spin around the Earth with much greater velocity? Why?
2. Does the honey propagate the sound? Explain.
3. We have situation as shown in the picture. Which listener will first hear the sound when the bell rings? Explain.

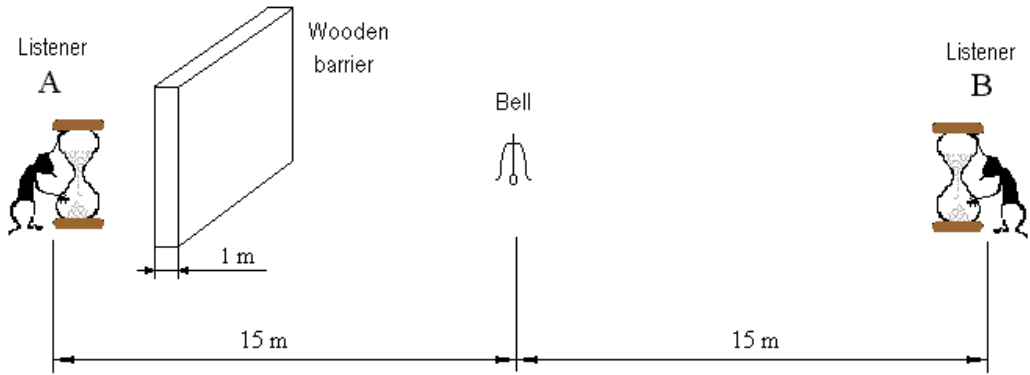

4. The ambulance car is going away from us with constant velocity. Its siren gives one tone. Will the pitch of this sound be changed if we start to run after the car? Explain.
5. Which guitar string gives the highest tone? Explain why?
6. A listener sits in the auditorium of an opera house If the management of theatre hang a big mirror (e.g. 2 x 2 m) from the roof so that the listener sees the singer on the stage in the mirror, would it influence the intensity of the sound which the listener receives from the singer? Explain.
7. Imagine the listener who is moving away at ultrasonic speed from the loudly playing orchestra. What does he hear from orchestra music? Explain. (Suppose that there is no noise in ears caused by velocity).
8. Does the sound of violin travel faster, slower or with the same velocity as the sound of contrabass, if we know that the violin produces higher frequencies. Explain.
9. Do compressions and rarefactions of the sound wave travel in the same or in opposite directions?
10. Does the sound of car siren travel equally fast when the car is at rest and when it is moving towards the observer?
11. Two tones are shown in pictures A and B (their shape is obtained by oscilloscope). Which of them is higher? Which is louder? How have you

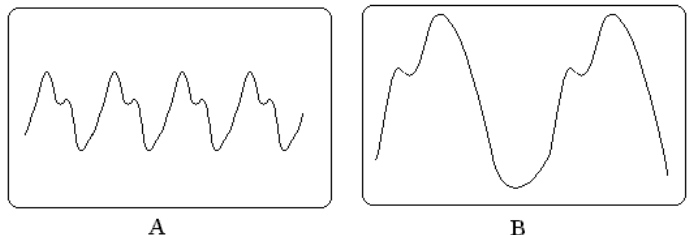

concluded it?

# GROUP D

1. Does the sound of strong loudspeaker travel faster, slower or with the same velocity as the sound of weak loudspeaker if they are both on maximum? Why?
2. The ambulance car is on plane platform a l00 m from us. We hear the sound of its siren. What change of the sound will occur if the car starts to circle around us with great velocity and at constant distance?

   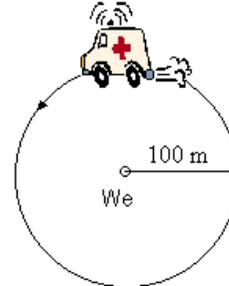

   Sketch of situation:

3. If we see an airplane in the sky, which flies with infrasonic velocity, will its sound come from the same spot on which we see the airplane? Explain.
4. What will change with guitar sound if we cover hole on its body? Why?
5. If the Earth revolution around its axis were much faster, would we hear the sound of this rotation? Why?
6. Does the plastics propagate sound? Explain.
7. Does the frequency, amplitude or both affect intensity of the sound? In which way? Explain this influence physically?
8. If we vibrate a guitar string, it stops after some time. Why?
9. Does the sound originate on the Moon when an astronaut jumps on it? Explain.
10. If we spread straw on the theatre stage, why does the choir give away less sound?
11. Imagine a wall completely impermeable for the sound built in open space. When watched from above it has the shape of the letter H, and it is 10 m high (see picture). We have two people there as shown at the picture. Can they hear each other? Explain why.

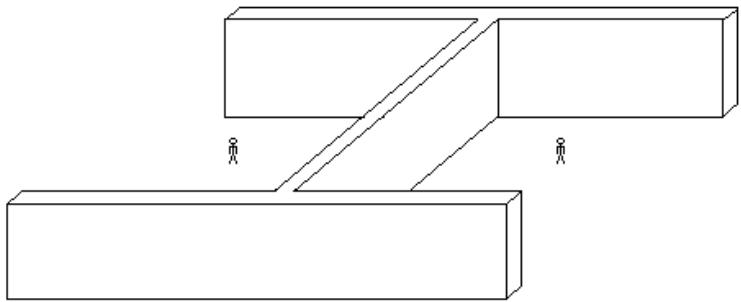